\documentclass[twocolumn]{aa}

\usepackage{graphicx}
\usepackage{txfonts}
\usepackage{natbib}
\newcommand{\beq}   {\begin{equation}}
\newcommand{\eeq}   {\end{equation}}
\newcommand{\kms}   {km~s$^{-1}$}
\newcommand{\water}   {H$_2$O~}

\begin{document}
   \title{Polarization of \water Masers in the presence of Velocity and Magnetic Field Gradients}

   \titlerunning{Polarization of \water Masers in the presence of Velocity and Magnetic Field Gradients}

   \subtitle{}

   \author{W.H.T. Vlemmings\inst{1}
          }

   \offprints{WV (wouter@jb.man.ac.uk)}

   \institute{Jodrell Bank Observatory, University of Manchester, Macclesfield,
                    Cheshire, SK11 9DL, England  
                          }

   \date{12-09-2005, A\&A accepted}

   \abstract{ Through polarization observations \water masers
   are excellent probes of magnetic fields in the maser
   region. Magnetic field strengths, such as those in the
   \water masers regions of the envelopes of late-type stars and star-forming
   regions, are typically determined using a direct relation between
   the field strength and the observed circular polarization. Here it is
   shown that velocity and magnetic field gradients along the maser
   have a significant effect on the field strengths obtained from
   circular polarization observations. Due to velocity gradients the
   actual magnetic field strength could be up to $100\%$ higher than
   the field strength derived from the observations. Additionally,
   when a magnetic field gradient is present, the resulting circular
   polarization derived is caused predominantly by the average
   magnetic field in the
   unsaturated maser core. Measurements of the fractional linear
   polarization are not affected by velocity or magnetic field
   strength gradients, though changes in the magnetic field angle
   along the maser do quench the linear polarization intensity when
   the maser saturates.

  }

   \maketitle

\section{Introduction}
Astrophysical \water masers, like other maser species such as OH
and SiO, are excellent probes of magnetic fields in a variety of
interesting regions. The magnetic field strength is derived from
observations of maser polarization caused by Zeeman
splitting. Polarization observations of the 22~GHz \water masers in the
circumstellar envelopes (CSEs) around evolved stars have made important
contributions to the understanding of the magnetic fields of evolved
stars \citep[][hereafter V02 and V05a]{V02,V05a}. \water maser observations have also provided
information on the strength and structure of the magnetic field in
star-forming regions \citep{FG89, S01, S02} and have provided upper limits for the field in the megamaser galaxy NGC 4258 \citep{M05}.

However, the exact relationship between the observed maser
polarization and the magnetic field strength is not straightforward,
leading to uncertainties in the determined field strength. It was
shown in \citet[][hereafter NW92]{NW92} that the influence of the
different hyperfine components of the \water ($6_{16}-5_{23}$)
rotational transition as well as of the degree of maser saturation is
significant. The effect of maser saturation on the magnetic field
strength determination can be as large as a factor of 2 (NW92).
Additionally, velocity and magnetic field gradients might play an
important role. While typical \water maser spot sizes are between
$\sim$10$^{12}$ and $10^{13}$~cm \citep{RM81, I97}, the actual size of
the maser region will be several times larger due to beaming
\citep[e.g.][]{GK72}. As a result the maser path length could be as
long as $10^{14}-10^{15}$~cm. Across such a path length, velocity and
magnetic field gradients likely influence the observed
polarization. In \citet[][hereafter V05b]{V05b} it was shown that
velocity gradients alter the shape of the total intensity line
profile. From this analysis and from earlier observations
\citep[e.g.][]{R99} it is apparent that in CSEs velocity shifts of
$\sim$1~\kms are common. In this paper the effects of velocity and
magnetic field gradients along the maser path on the magnetic field
strength determined from circular polarization measurements and on the
observed fractional linear polarization are examined.

\begin{figure*}[ht!]
   \resizebox{0.9\hsize}{!}{\includegraphics{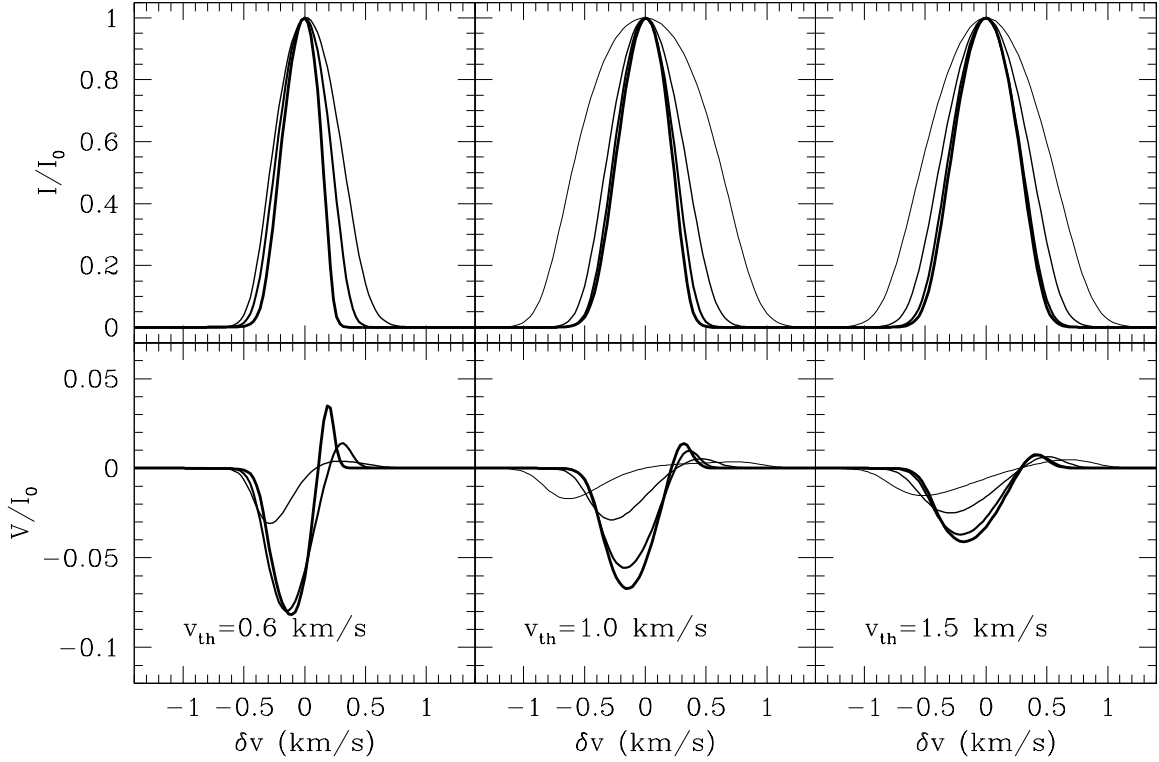}}
   \hfill
\caption[]{Normalized total intensity (I) and circular polarization (V) spectra for $v_{\rm th}=0.6, 1.0$ and $1.5$~\kms in the presence of a velocity gradient. From thick to thin the solid lines denote velocity shifts of $\Delta V_m = 0, 0.5, 1.0$ and $1.5$~\kms respectively except for $v_{\rm th}=0.6$~\kms where $\Delta V_m = 1.5$~\kms is omitted.}
\label{Fig:ex}
\end{figure*}

\section{Background}
\label{back}

Similar to the analysis in V02 and V05b, the equations of state for
the populations of the upper ($6_{\rm 16}$) and lower ($5_{\rm 23}$)
rotational levels of the three strongest hyperfine components
($F-F'=7-6, 6-5$ and $5-4$) of the 22~GHz \water maser have been
solved using the method described in NW92. The equation of state
for the number density $n(F,a,\nu)$ of the upper energy levels ($F$)
is

\begin{eqnarray}
0 & = & \lambda_F(v_s) - (\Gamma+\Gamma_v)n(F,a,v) \nonumber \\
& & + {R(F,F',a,b,v)}(n(F',b,v) - n(F,a,v)) \nonumber \\ 
& & + \phi(v_s)({\Gamma_v \over \sum g_{F}})\int dv\sum g_{F} n(F,a,v).
\label{eq1}
\end{eqnarray}

The equation for the number density $n(F',b,\nu)$ of the lower levels
$(F')$ is similar but reversed. Here $a$ and $b$ denote the different
magnetic sub-states of each hyperfine component. The population levels
are solved as a function of molecular velocity $v$ and at different
positions along the maser propagation direction. The statistical
weights are designated by $g_F$ and $g_F'$, and are from
\citet{K69}. The pump rate $\lambda_F(v_s)$ is assumed to be the same
for the different hyperfine components and has a Maxwellian
distribution.  The results of the calculations depend only on the the
ratio $(\lambda_F - \lambda_F') / \lambda_F$, which is of the order of
a few percent \citep{AW93}. Generally $v_s=v$, but when velocity
gradients are introduced later, the molecular velocity of the maser
pump $v_s$ changes along the maser path. The rate for stimulated
emission $R(F,F',a,b,v)$ (hereafter $R$) is calculated using the local
maser intensity and the hyperfine interaction coefficients as
described in NW92. $\Gamma$ is the decay rate for the molecular
excitations.  The cross-relaxation rate $\Gamma_\nu$ describes the
reabsorbtion of previously emitted infrared pump photons trapped in
optically thick transitions of the maser system, which generate a
newly excited molecule at random, within a Maxwellian velocity
distribution ($\phi(v_s)$). 
Elastic collisions between the maser molecules and intermixed H$_2$
molecules have not been included. The effects of these are described
in detail in \citet{E90} and are unimportant for the \water masers
discussed here.

\begin{figure*}[t!]
   \resizebox{0.9\hsize}{!}{\includegraphics{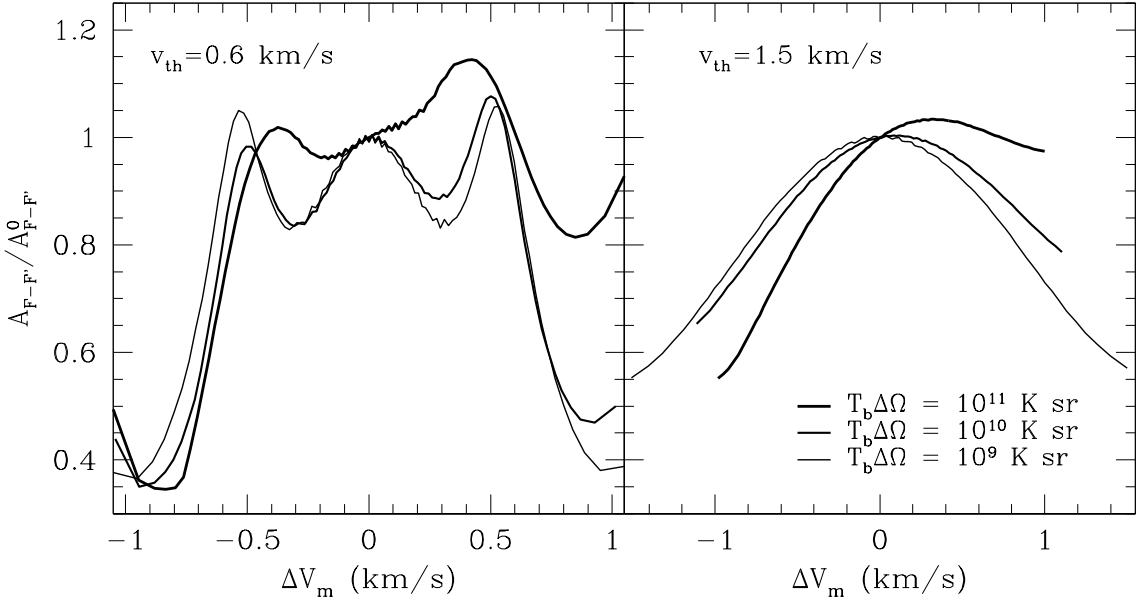}}
   \hfill
\caption[]{The dependence of the magnetic field strength determined from circular polarization observation, represented by the $A_{F-F'}$ coefficient from Eq.~\ref{eq2}, on the magnitude of the velocity gradient $\Delta V_m$ for two values of intrinsic thermal line width $v_{\rm th}$. $A_{F-F'}$ is normalized by $A^0_{F-F'}$, which is $A_{F-F'}$ for $\Delta V_m=0$~\kms. The solid lines represents masers with different emerging brightness temperatures. Note that the noise ($<1$\%) on the curves is due to the velocity resolution of our model spectra. The results are independent of the magnetic field angle $\theta$.}
\label{affvsv}
\end{figure*}


The maser intensity, which is solved using the radiative transfer
equations from NW92, influences the level populations through
$R$. However, as it is itself dependent on the level populations
determined with Eq.~\ref{eq1}, the maser total intensity (I), linear
polarization (Q) and circular polarization (V) are solved iteratively
along the maser path. Assuming nearly one-dimensional maser
propagation the beaming of the maser radiation is represented by the
solid angle $\Delta\Omega$. The emerging maser fluxes will thus be
represented in $T_{\rm b}\Delta\Omega$, with $T_{\rm b}$ the
brightness temperature. All calculations were performed with a
brightness temperature for the radiation incident on the maser region
of $T_{\rm b}\Delta\Omega = 0.1$ K~sr. It was verified in NW92 and V02
that the results are insensitive to the chosen initial value. The
calculations are performed for different thermal line widths ($v_{\rm
th}$) of the Maxwellian particle velocity distribution, where $v_{\rm th}\!\approx$0.5$ (T/100)^{1/2}$~\kms with $T$ the temperature of the masing
molecules in $K$. 
\citet{NW91} and NW92 have shown that when Eq.~\ref{eq1} is
solved without including the contribution of the cross-relaxation rate
$\Gamma_\nu$, the results for the line profiles, fractional
polarizations and line width in the case of non-negligable
$\Gamma_\nu$ can be obtained by scaling the results with
$[\Gamma+\Gamma_\nu]$. Following this result, the calculations in this
paper are performed for $\Gamma=1$~s$^{-1}$ and
$\Gamma_\nu=0$~s$^{-1}$. Thus, in all the results of this
paper, the emerging brightness temperature $T_{\rm b}\Delta\Omega$ can
be replaced with $T_{\rm b}\Delta\Omega \times [\Gamma+\Gamma_\nu]$ to
obtain the results for different values of the decay and
cross-relaxation rates.

To solve the equations of state it is assumed that the Zeeman
frequency shift $g\Omega$ is much greater than $R$, $\Gamma$ and
$\Gamma_\nu$. Then the off-diagonal elements of the density matrix
describing the molecular states are negligible, greatly simplifying
the calculations. For a magnetic field strength of $\sim$1.0~G,
$g\Omega$ of the dominant hyperfine transitions is
$\sim$10$^4$~s$^{-1}$ while for $T_{\rm b}\Delta\Omega < 10^{12}$~K~sr
the rate for stimulated emission $R \leq 100$~s$^{-1}$ (V02). Most of
the calculations in this paper are performed for a magnetic field
strength of $B=1$~G. For the highest $T_{\rm b}\Delta\Omega$, the
results will be valid for fields down to $\sim$10~mG, when $g\Omega$
becomes comparable to $R$. However, the low observed linear polarization of the 22~GHz \water masers (e.g. V02, V05a) indicates that
$T_{\rm b}\Delta\Omega$ is at most $10^{11}$~K~sr and typically even lower. Thus, the results presented in this paper are valid for $B$ well below $1$~mG.

22~GHz \water masers begin to saturate when $R/\Gamma \gtrsim
  1$~s$^{-1}$ and approach full saturation when $R/\Gamma \gtrsim
  100$~s$^{-1}$. Thus, for our models with $\Gamma=1$~s$^{-1}$,
saturation starts at $T_{\rm b}\Delta\Omega \gtrsim 10^{10}$~K~sr and
full saturation occurs for $T_{\rm b}\Delta\Omega \gtrsim
10^{12}$~K~sr. The maser spectral line profile (through rebroadening)
and the polarization properties of the maser changes at
$R=(\Gamma+\Gamma_\nu)$ (e.g. NW92). While typically for the 22~GHz
\water masers this is close to when saturation occurs, a large
cross-relaxation rate means the changes will only occur when the maser
has already become partly saturated. As $\Gamma_\nu$ for the
  astrophysical \water masers is at most $\approx 5$~s$^{-1}$ for
  \water masers at $T\sim 1000~K$ and is less than that lower
  temperatures \citep{AW93}, $\Gamma_\nu\approx\Gamma$ is assumed in
  this paper when discussing unsaturated and saturated masers. It is
  worth noting that \water maser observations indicate that in
  actuality $\Gamma$ is between one and two orders of magnitude less
  than $\Gamma_\nu$ \citep{AW93}. However, as our results scale
  linearly with $(\Gamma+\Gamma_\nu)$ this does not affect the
  conclusions in this paper.

\begin{figure*}[t!]
   \resizebox{0.9\hsize}{!}{\includegraphics{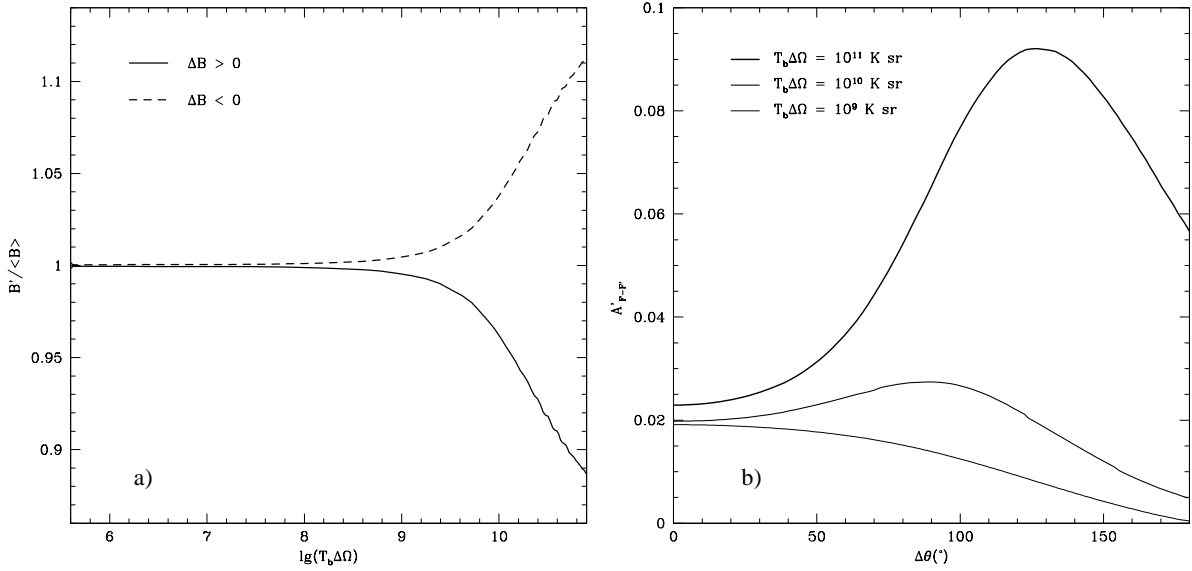}}
   \hfill
\caption[]{a) Derived magnetic field strength $B'$ over the average field strength along the maser $\langle B\rangle$ as a function of $T_{\rm b}\Delta\Omega$. The solid and dashed lines correspond to positive and negative magnetic field gradients respectively. b) $A'_{F-F'}$ as a function of $\Delta\theta$ for different values of emerging brightness temperature $T_{\rm b}\Delta\Omega$. Calculations were performed for $v_{\rm th}=0.8$~\kms.}
\label{bchange}
\end{figure*}

\section{Radiative Transfer Modeling Results}

\subsection{Circular Polarization}

As shown in NW92 and V02, the V-spectra produced from the radiative
transfer equations are not anti-symmetric. They are also not always
directly proportional to the derivative of the I-spectrum due to
hyperfine interaction as is often assumed. However, as was discussed
in V02, it is impossible to directly observe the intrinsic shape of
the V-spectrum due to instrumental effects and necessary data
calibration steps. Still, the percentage of circular polarization
$P_{\rm V}$ can be related to the magnetic field strength using
\begin{eqnarray}
P_{\rm V} & = & (V_{\rm max} - V_{\rm min})/I_{\rm max} \nonumber\\
& = & 2\cdot A_{F-F'}\cdot B_{\rm [Gauss]} \rm{cos}\theta/\Delta v_{\rm L}[\rm{km~s^{-1}}].
\label{eq2}
\end{eqnarray}
Here $V_{\rm max}$ and $V_{\rm min}$ are the minimum and maximum of
the V-spectrum. $I_{\rm max}$ and $\Delta v_{\rm L}$ are the peak flux
and the full width half-maximum (FWHM) of the I-spectrum respectively.
$B$ is the magnetic field strength at an angle $\theta$ to the maser
propagation direction. While in most analyses of maser circular
polarization, the coefficient $A_{F-F'}$ is taken to be a fixed value
\citep[e.g][]{FG89}, it was shown in NW92 to depend on the level of
maser saturation and the intrinsic thermal velocity $v_{\rm th}$ of
the maser. Figure 6 of V02 shows $A_{F-F'}$ for different values of
$v_{\rm th}$ as a function of emerging maser brightness temperature
$T_{\rm b}\Delta\Omega$. 
For maser brightness temperatures $T_{\rm b}\Delta\Omega >
10^9$~K~sr it was shown in NW92 that the ${\rm cos}\theta$ dependence
of Eq.~\ref{eq2} breaks down effectively also introducing a dependence
on $\theta$ to $A_{F-F'}$. This was later shown in more detail in
\citet{WW01} for masing involving angular momentum $J=$1--0 and
$J=$2--1 transitions. In V02, figure 7 shows the derived magnetic
field strength dependence on $\theta$ to $A_{F-F'}$ for the 22~GHz
$J=$6--5 transition.

\subsubsection{velocity gradients}

\begin{figure*}
   \resizebox{0.9\hsize}{!}{\includegraphics{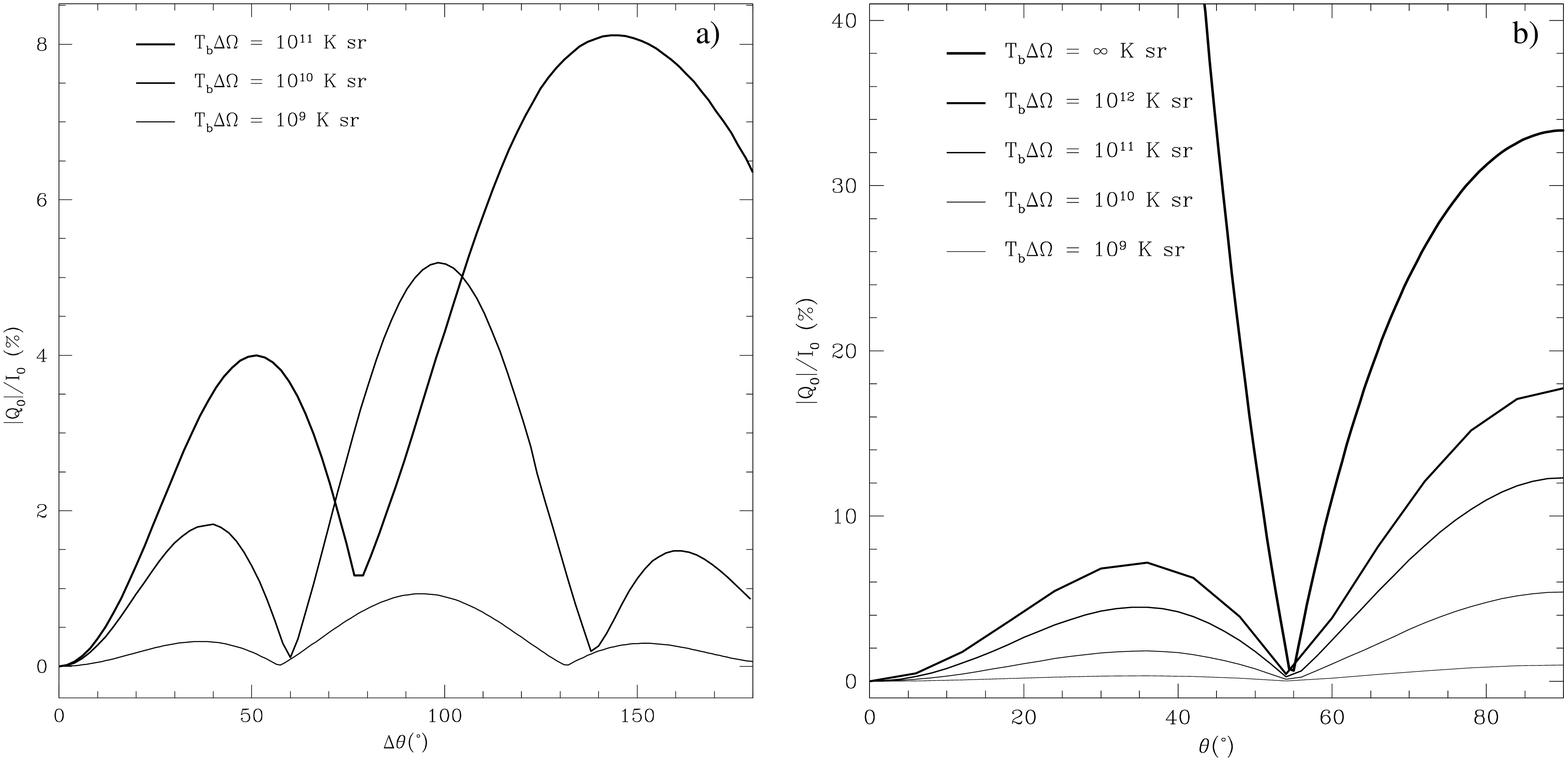}}
   \hfill
\caption[Linpol]{a) Magnetic field angle gradient $\Delta\theta$ vs. the fractional linear polarization for three different stages of maser saturation and $\theta_0 = 0^\circ$. b) The angle $\theta$ between the maser propagation direction and the magnetic field vs. the fractional linear polarization for different values of emerging maser brightness. The thick solid line denotes the theoretical limit from \citet{GKK} for a completely saturated maser.}
\label{Fig:linpol}
\end{figure*}

First, a velocity gradient is introduced by uniformly shifting, in
Eq.~\ref{eq1}, the velocity $v_s$ of the pump rate $\lambda_F$ for
each integration step along the maser path. By varying the amount of
shift, a velocity difference $\Delta V_m$ is created between the start
and end of the maser amplification path of up to $\sim$1.5~\kms. This
corresponds to a velocity gradient of $\Delta V_m / S$~\kms m$^{-1}$,
where $S$ is the length of the maser path. As $S$ depends on the exact
values of the pumping rate $\lambda_F$, while our results, expressed
in emerging brightness temperature are only dependent on the ratio of
pumping rates as addressed above, any mention of the velocity gradient
will henceforth be referring to the value of $\Delta V_m$. In
\citet{NW88} it was shown that the maser splits into several
distinctly separate narrow features when $\Delta V_m$ becomes a few
times $v_{\rm th}$, each of the features becoming an independent
maser. The calculations in this paper are therefore limited to values
of $\Delta V_m$ up to $\sim$1.5~\kms.

Fig.~\ref{Fig:ex} shows the effect of a velocity gradient on the total
intensity and circular polarization profiles for three different
intrinsic thermal velocities $v_{\rm th}$, produced for the same
magnetic field strength (1~G). While for low $\Delta V_m$, the effect
of the velocity gradient on the shape of the I-profile is small, the
effect on the V-spectrum can still clearly be seen. To examine how a
velocity gradient influences the magnetic field determination from the
circular polarization measurements, the normalized $A_{F-F'}$
coefficient from Eq.~\ref{eq2} is plotted in Fig.~\ref{affvsv} as a
function of velocity gradient for $v_{\rm th}=0.6$ and $1.5$~\kms and
emerging brightness temperatures $T_{\rm b}\Delta\Omega = 10^9,
10^{10}$ and $10^{11}$~K~sr. Because of the normalization, the
results are independent of the magnetic field angle $\theta$. For
$v_{\rm th}=0.6$~\kms the effect of the hyperfine components is
clear. For $T_{\rm b}\Delta\Omega = 10^9$~K~sr none of the hyperfine
components are becoming saturated yet, thus $A_{F-F'}$ is mostly
symmetric around $\Delta V_m = 0$~\kms. The symmetry breaks down when
the hyperfine lines are slowly starting to saturate for $T_{\rm
b}\Delta\Omega \gtrsim$10$^{10}$~K~sr, as maser growth is somewhat
inhibited for velocity gradients in the direction of the weaker
hyperfine components ($F-F'=6-5$ and $5-4$), while a gradient in the
opposite direction allows for stronger maser amplification. As
noted above this effect can be postponed further into the saturated
regime when $\Gamma_\nu$ increased. It can be seen in the figure, that
$A_{F-F'}$ typically decreases for increasing $|\Delta V_m|$, implying
that in the presence of velocity gradients the magnetic field strength
$B'$ derived with Eq.~\ref{eq2} underestimates the true $B$. For small
$v_{\rm th}$ the field strength can sometimes be underestimated by
more than a factor of $2$. However, for a narrow range in $\Delta
V_m$, $A_{F-F'}$ is actually enhanced. In those cases $B$ could
actually be overestimated by up to $\sim$20\%. This effect is largest
for the more saturated masers.

\subsubsection{magnetic field gradients}

There are different ways a magnetic field gradient along the maser
amplification path can occur. The actual field strength $B$ can change
or the angle $\theta$ between the field and the maser propagation
direction can vary. Similar to $\Delta V_m$, differences
$\Delta\theta$ and $\Delta B$ are introduced between the start and end
of the maser path and their influence on the maser polarization is
examined. Fig.~\ref{bchange}a shows the ratio between the magnetic
field $B'$ determined with Eq.~\ref{eq2} and the average magnetic
field $\langle B\rangle$ along the maser as a function of $T_{\rm
b}\Delta\Omega$ when a gradient $\Delta B$ is included. The shape of
this function does not depend on the actual value of $\Delta B$ or on
the intrinsic thermal line width $v_{\rm th}$. While the maser is
unsaturated, or more precisely while $R<\Gamma_\nu$, $B'$ is
equal to $\langle B\rangle$. However, when the maser
saturates, $B'$ will be dominated by the average magnetic field in the
unsaturated maser region.

Fig.~\ref{bchange}b shows the change in $A'_{F-F'}$ versus
$\Delta\theta$ for different $T_{\rm b}\Delta\Omega$. The angle
$\theta_0$, the value of $\theta$ at the start of maser amplification
is taken to be $0^\circ$. Here $A'_{F-F'} = A_{F-F'}$cos$\theta$, as
the actual angle $\theta$ cannot be specified due to the inclusion of
the gradient $\Delta\theta$. Comparing the unsaturated maser ($T_{\rm
b}\Delta\Omega = 10^9$~K~sr) with the partly saturated maser ($T_{\rm
b}\Delta\Omega = 10^{11}$~K~sr) it is again apparent that, as seen in
Fig.~\ref{bchange}a, the unsaturated maser regime provides the
dominant contribution to the magnetic field strength determined with
Eq.~\ref{eq2}. Comparing Fig.~\ref{bchange}b with figure 7 from V02
indicates that the angle average $\langle\theta\rangle$ in the
unsaturated maser region determines the observed circular
polarization.

\subsection{Linear Polarization}

The linear polarization in the presence of a magnetic field was also
discussed in NW92. The fractional linear polarization is not found to
be affected when including a velocity gradient $\Delta V_m$ or
magnetic field gradient $\Delta B$ in the calculations. Only when
including a gradient $\Delta\theta$ does the linear polarization
change. In Fig.~\ref{Fig:linpol}a the fractional linear polarization
$|Q_0|/I_0$ is shown as a function of $\Delta\theta$.  This can be
compared with $|Q_0|/I_0$ as a function of the angle between the
magnetic field and the maser propagation direction $\theta$ shown in
Fig.~\ref{Fig:linpol}b. This relationship, in the limiting case of a
completely saturated maser, was solved in \citet{GKK}. In the case of
an unsaturated maser ($T_{\rm b}\Delta\Omega \lesssim 10^{10}$~K~sr), the
relationship between $|Q_0|/I_0$ and the magnetic field angle gradient
$\Delta\theta$ is very similar to the relationship between $|Q_0|/I_0$
and $\theta$, while when the saturation level increases the
linear polarization is quenched and shifted. This indicates that the
observed fractional linear polarization is determined by $\theta$ at
the end of the unsaturated maser regime, where the largest
amplification occurs, and not by the average $\langle\theta\rangle$
over the unsaturated maser core.

\section{Conclusions}
\label{concl}

Using a maser radiative transfer code which includes the three
strongest hyperfine components of the 22~GHz \water $6_{\rm 16} -
5_{\rm 23}$ rotational transition and their magnetic sub-states, the
effects of velocity and magnetic field gradients on the determination
of magnetic fields from maser polarization observations have been
calculated.  It was shown that, due to velocity gradients, the
magnetic field strength determined on \water masers in CSEs (V02,
V05a), where $v_{\rm th}\!\approx$0.8$-1.0$~\kms and where velocity
gradients of $\sim$1.0~\kms have been observed (V05b), can be
underestimated by up to a factor of 2. However, for maser features
with smaller velocity gradients there will be cases where $B$ is
actually overestimated by $\sim$10$-20\%$. The magnetic fields
determined on the \water masers in star-formation regions \citep{FG89,
S01, S02}, which have higher thermal line widths ($v_{\rm
th}\!\approx$1.5~\kms) can be underestimated by $\sim$20$-40\%$. These
uncertainties can only be overcome by including a consistent model for
the velocity gradients throughout the \water maser source when
deriving the magnetic field strengths derived from circular
polarization observations.

When there is a magnetic field gradients along the maser amplification
path, $B'$ determined from polarization observations of an unsaturated
maser (more formally when $R\lesssim\Gamma_\nu$), is typically
equal to the average field along the maser. When the maser is
saturated, the $B'$ is dominated by the average magnetic field
strength or the average magnetic field angle in the unsaturated maser
core. The fractional linear polarization on the other hand, is not
affected by velocity or magnetic field gradients. Only a gradient in
the angle between the magnetic field and the maser propagation axis
alters the linear polarization. In contrast to the circular
polarization, the fractional linear polarization is mainly determined
by the magnetic field angle in the part of the unsaturated maser where
the largest amplification occurs and not by the angle averaged over
the entire unsaturated maser region.

\begin{acknowledgements}
The author acknowledges the helpfull comments by the anonymous
referee.  This work has been supported by an EC Marie Curie Fellowship
under contract number MEIF-CT-2005-010393.  
\end{acknowledgements}

\bibliographystyle{aa}

\end{document}